\documentclass[12pt,aps,prd,preprint,tightenlines,superscriptaddress,showpacs,nofootinbib]
{revtex4-1}
\usepackage{epsfig}
\usepackage{amssymb,amsmath}
\usepackage{hyperref}
\usepackage{xfrac}
\usepackage{color}
\usepackage[utf8]{inputenc} 
\usepackage{feynmp-auto}

\newcommand{\bea}{\begin{eqnarray}}
\newcommand{\eea}{\end{eqnarray}}
\begin{document}
%\begin{flushright}
%\end{flushright}
%
\vspace*{1.0cm}

\begin{center}
\baselineskip 20pt 
{\Large\bf 
Nambu-Goldstone boson phenomenology \\ in Domain-Wall Standard Model
}
\vspace{1cm}

{\large
Puja Das $^{a,}$\footnote{pdas1@crimson.ua.edu}, 
Nobuchika Okada $^{a,}$\footnote{okadan@ua.edu}, 
Digesh Raut $^{b,}$\footnote{draut@smcm.edu}, and 
Desmond Villalba $^{c,}$\footnote{dvillalb@umw.edu} 
}
\vspace{.5cm}

{\baselineskip 20pt \it
$^a$Department of Physics and Astronomy, \\University of Alabama, Tuscaloosa, AL 35487, USA\\
$^b$Physics Department, St.~Mary's College of Maryland, St. Mary's City, MD 20686, USA\\
$^c$Department of Chemistry and Physics, \\University of Mary Washington, Fredericksburg , VA 22401, USA
} 

\vspace{.5cm}

\vspace{1.5cm} {\bf Abstract}
\end{center}

We investigate the Domain-Wall Standard Model (DWSM), a five-dimensional framework in which all Standard Model (SM) particles are localized 
  on a domain wall embedded in a non-compact extra spatial dimension. 
A distinctive feature of this setup is the emergence of a Nambu-Goldstone (NG) boson, 
  arising from the spontaneous breaking of translational invariance in the extra dimension due to the localization of SM chiral fermions.
This NG boson couples via Yukawa interactions to SM fermions and their Kaluza-Klein (KK) excitations.
We study the phenomenology of this NG boson and derive constraints from astrophysical processes (supernova cooling), 
  Big Bang Nucleosynthesis (BBN), and collider searches for KK-mode fermions at the Large Hadron Collider (LHC).
The strongest limits arise from LHC data: we reinterpret existing mass bounds on squarks and sleptons in simplified supersymmetric models 
  (assuming a massless lightest neutralino), as well as limits on exotic hadrons containing long-lived squarks or long-lived charged sleptons 
  in the regime of extremely small Yukawa couplings.
From this analysis, we obtain a conservative lower bound of 1 TeV on the masses of KK-mode quarks and charged leptons.
Finally, we discuss the prospects for producing KK-mode fermions at future high-energy lepton colliders and outline strategies 
  to distinguish their signatures from those of sfermions.

\thispagestyle{empty}

%\bigskip
\newpage

\addtocounter{page}{-1}

%%%%%%%%%%%%%%%%%%%%%%%%%%
%\baselineskip 36pt
% Main body
%%%%%%%%%%%%%%%%%%%%%%%%%%
%\baselineskip 18pt
%%%%%%%%%%%%%%%%%%%%%%%%%%

%%%%%%%%%%%%%%%%%%%%%%%%
\section{Introduction} 
\label{sec:1}
%%%%%%%%%%%%%%%%%%%%%%%%

The possibility that our universe consists of more than three-dimensional space has long attracted interest, 
   dating back to the original idea of unifying gravity with electromagnetism proposed by Einstein, Kaluza, 
   and Klein nearly a century ago.
Following the discovery of D-branes in string theory \cite{Polchinski:1995mt}, their phenomenological applications 
   to physics beyond the Standard Model (SM), the so-called braneworld scenario, have been extensively studied. 
Two well-known realizations are the large extra-dimension model \cite{Arkani-Hamed:1998jmv, Antoniadis:1998ig} 
   and the warped extra-dimension model \cite{Randall:1999ee}, 
   both of which offer solutions to the gauge hierarchy problem alternative to supersymmetric extensions of the SM.  
  
In the braneworld scenario, the extra spatial dimensions are typically compactified on specific manifolds and/or orbifolds, 
   and treated differently from our three spatial dimensions. 
The four-dimensional effective theory is obtained by integrating out the finite volume of the extra dimensions, 
   where massive Kaluza-Klein (KK) modes emerge and lead to interesting phenomenology. 

However, it is arguably more natural to consider scenarios where the extra spatial dimensions are non-compact, 
   just like our familiar three dimensions. 
In such a case, a viable four-dimensional effective theory requires all SM fields as well as the graviton to be localized 
   within a three-dimensional subspace (domain) embedded in the higher-dimensional bulk.
Randall and Sundrum \cite{Randall:1999vf} showed that in a five-dimensional gravity theory with anti-de Sitter (AdS) curvature, 
   the graviton zero-mode can be localized near a point along the non-compact fifth dimension. 
In this setup, all SM fields are confined to a 3-brane located at that point, 
   leading to a low-energy effective theory consistent with Einstein gravity and the SM in four dimensional spacetime.  
Cosmological applications of this framework have been studied, 
   and remarkably, viable solutions have been found \cite{Binetruy:1999ut, Binetruy:1999hy, Shiromizu:1999wj, Ida:1999ui}, 
   in which the early universe cosmology asymptotically approaches the standard four-dimensional Big Bang cosmology
   as the SM plasma cools during cosmic expansion.

Some of the present authors \cite{Okada:2017omx, Okada:2018von} and another group \cite{Arai:2018uoy} 
   have proposed a framework that we call the ``Domain-Wall Standard Model", formulated in a non-compact five-dimensional spacetime. 
In this model, all SM fields are localized within finite-width regions (domain walls) along the fifth dimension, 
   providing a field-theoretic realization of 3-brane. 
 A key difference from 3-brane is that the domain walls possess finite thickness. 
This localization mechanism gives rise to KK modes, massive partners of SM particles, with their masses characterized 
  by the domain-wall width, 
The KK modes offer rich phenomenology, 
  including possible signals at the CERN Large Hadron Collider (LHC) \cite{Okada:2017omx, Okada:2018von} 
  and implications for flavor physics \cite{Okada:2019fgm}. 
 For earlier works discussing gauge theory localized on walls in higher dimensional spaces, see Refs.~\cite{Davies:2007xr, Arai:2012cx, Arai:2013mwa, Arai:2017ntb, Arai:2017lfv, Arai:2021wul}.
  
The localization mechanism varies depending on the type of field, and we briefly summarize below 
  the key ideas that are generally applicable.  
\begin{enumerate}
\item Domain-Wall Gauge Fields \\
Localization of gauge bosons is achieved by introducing a five-dimensional gauge Lagrangian 
  with a gauge coupling $g(y)$ that depends on the extra-dimensional coordinate \cite{Ohta:2010fu} 
  (see, also, \cite{Luty:2002hj}).
This approach offers a simpler alternative to earlier non-perturbative localization mechanisms for gauge fields \cite{Dvali:1996xe}.
The formalism has been refined in Refs.~\cite{Okada:2017omx, Okada:2018von, Arai:2018rwf}, 
  incorporating a gauge-fixing term analogous to the $R_\xi$-gauge in four-dimensional gauge theories.  
For a few choices of $g(y)$, exact solutions for the KK mode equations can be obtained, 
  yielding the full KK-mode spectrum and corresponding mode functions. 

\item Domain-Wall Higgs Field \\
As shown in \cite{Okada:2017omx, Okada:2018von}, the Higgs field can be localized using a profile function 
  $s_H(y)$, such that the five-dimensional Higgs Lagrangian is proportional to $s_H(y)$.
By choosing $s_H(y) \propto 1/g(y)^2$, the Higgs mechanism operates in the same way as in the four-dimensional theory 
  for zero-mode gauge bosons, while the squared masses for their corresponding KK-modes 
  are shifted by the zero-mode squared masses generated by the Higgs  mechanism. 

\item Domain-Wall Fermions \\
For fermion localization, we follow the mechanism proposed by Rubakov and Shaposhnikov \cite{Rubakov:1983bb}.
A real scalar field with a double-well potential is introduced in the five-dimensional sapce, 
  along with a bulk fermion that couples to it via a Yukawa interaction. 
The scalar field develops a kink solution along the extra-spatial dimension. 
Solving the fermion equations of motion, we find that one chiral component is localized at the kink center while the other is delocalized. 
The resulting four-dimensional effective theory thus contains a chiral fermion, accompanied by KK excitations. 
Interestingly, depending on the parameters of the scalar potential and the Yukawa coupling, 
  one can realize KK-mode spectrum with a finite number of discrete mass eigenstates.
\end{enumerate}

Under the kink background, which is essential for realizing domain-wall fermions, 
  the equation of motion for a dynamical scalar field leads to two localized modes: one massless and one massive. 
Note that the massless mode is a Nambu-Goldstone (NG) boson 
  associated with the spontaneous breaking of translational invariance in the extra-spatial dimension.
The four-dimensional effective Lagrangian includes Yukawa interactions 
  among this NG boson, the SM chiral fermions and their KK-mode partners, 
  inherited from the original five-dimensional Yukawa terms.
The presence of this NG boson distinguishes the Domain-Wall Standard Model 
   from the braneworld scenarios with compactified extra dimensions. 
In compactified models, bulk fermions also give rise to KK modes in the effective theory, but the NG boson is absent 
  due to the explicit breaking of translational invariance due to compactification. 
In contrast, our model retains this symmetry at the Lagrangian level, and its spontaneous breaking leads to observable effects.

In this paper, we investigate the phenomenology of the NG boson in the Domain-Wall Standard Model.
The NG boson has Yukawa couplings with the SM fermions and their KK excitations, allowing processes such as the decay 
  of a KK-mode fermion into an SM fermion and an NG boson, as well as SM fermion pair annihilation into NG bosons, and vice versa.
These interactions can occur in a variety of environments, including supernova cores, the early universe, and high-energy colliders.
We analyze these processes and derive constraints on the NG boson's Yukawa couplings and KK-mode fermion masses.
For simplicity, we assume that the KK excitations of the SM gauge and Higgs bosons are sufficiently heavy and 
   thus do not significantly affect the NG boson phenomenology considered in this work.

This paper is organized as follows: 
In the next section, we present the basic formalism of domain-wall fermions and derive the Yukawa couplings 
  of the NG boson in the four-dimensional effective theory. 
Section~\ref{sec:3} is devoted to phenomenological constraints from supernova cooling, Big Bang Nucleosynthesis (BBN), 
   and searches for KK-mode quarks and leptons at the LHC.  
The last section is devoted to conclusion and discussion.

%%%%%%%%%%%%%%%%%%%%%%%%%%%%%%%%%%%%%%%%%%%%%%%%%%%%%%%%%%
\section{Domain-Wall fermions and NG-mode coupling}
\label{sec:2}
%%%%%%%%%%%%%%%%%%%%%%%%%%%%%%%%%%%%%%%%%%%%%%%%%%%%%%%%%%
In this section, we derive the formulas necessary for the analysis in the next section. 
As mentioned above, we follow the mechanism proposed in Ref.~\cite{Rubakov:1983bb} to obtain a chiral gauge theory, 
  the Domain-Wall SM, in the four-dimensional effective theory.

%%%%%%%%%%%%%%%%%%%%%%%%%%%%%%
\subsection{Kink solution along the extra spacial direction}
\label{sec:2-A}
%%%%%%%%%%%%%%%%%%%%%%%%%%%%%%
We begin with a real scalar field ($\varphi(x, y)$) in the bulk: 
\begin{eqnarray} 
 {\cal L}_{5} = 
  \frac{1}{2} \left(\partial_{M} \varphi \right) \left( \partial^{M} \varphi \right) - V(\varphi)   \; ,  
\end{eqnarray}
where $M=0,1,2,3,y$ with $y$ being the index for the fifth-dimensional coordinate, and the scalar potential is given by 
\bea 
 V(\varphi) = \frac{m_\varphi^4}{2 \lambda} 
	   -m_\varphi^2 \varphi^2  
           + \frac{\lambda}{2} \varphi^4 .  
\eea
Throughout this paper, we use $x^\mu$ ($\mu=0, 1, 2, 3)$ for the conventional four-dimensional spacetime coordinates 
  while  $y$ for the fifth-dimensional coordinate, with the metric $\eta_{M N}=diag(1,-1,-1,-1,-1)$. 

It is well known that the equation of motion admits a non-trivial background configuration $\varphi_0 (y)$, 
  the so-called kink solution, given by
\bea
\varphi_{\rm kink} (y) = \frac{m_\varphi}{\sqrt{\lambda}} 
\tanh [m_\varphi y] , 
\label{kink} 
\eea
where we have chosen the kink center to be located at $y=0$.
Expanding the scalar around the kink background with the KK-mode decomposition, 
  $\varphi (x, y) = \varphi_{\rm kink}(y) + \sum_{n=0} \tilde{\chi}^{(n)}(y) \,  \varphi^{(n)}(x)$, 
  we obtain the linearized equation of motion for the $n$-th KK-mode: 
\bea
 \left[ -\partial_y^2 - \frac{\gamma (\gamma+1) m_\varphi^2}{\cosh^2(m_\varphi y)} \right] \tilde{\chi}^{(n)} 
  = \left(m_n^2 - \gamma^2 m_\varphi^2 \right)  \tilde{\chi}^{(n)}, 
\label{KK_EOM_phi}
\eea
where $\gamma=2$. 
This equation has the same form as the time-independent Schr\"odinger equation, $(-\partial_y^2 +V) \Psi^{(n)}=E_n \Psi^{(n)}$,  
   in one-dimensional quantum mechanics with a potential well, $V \propto -1/\cosh^2(m_V y) < 0$. 
Finding the states localized around the kink center is equivalent to finding bound states 
   in this quantum mechanics with $E_n <0$ that satisfy the boundary condition, $\tilde{\chi}^{(n)} \to 0$ for $|y| \to \infty$.
Solving this equation, we find two bound states, namely, one massless and the other massive:
\bea
   \varphi (x, y) = \varphi_{\rm kink}(y) +  \frac{\sqrt{3 m_\varphi} }{2} \left[ \frac{1}{\cosh^2(m_\varphi y)} \right] \varphi^{(0)}(x) 
    + \sqrt{ \frac{3m_\varphi}{2} } \left[ \frac{\sinh (m_\varphi y)}{\cosh^2 (m_\varphi y)} \right] \varphi^{(1)}(x), 
\label{s-mode}    
\eea  
where $\varphi^{(0)}(x)$ is a massless NG mode corresponding to the spontaneous breaking 
  of the translational invariance in the fifth dimension, 
  and $\varphi^{(1)}(x)$ is a massive mode with mass $m_\varphi^{(1)} = \sqrt{3} m_\varphi$.  
Here, the KK mode functions are normalized so that the kinetic terms of the corresponding four-dimensional eigenstates 
  are canonically normalized in the effective theory.

%%%%%%%%%%%%%%%%%%%%%%%%%%%%%%
\subsection{Domain-Wall Fermion}
\label{sec:2-B}
%%%%%%%%%%%%%%%%%%%%%%%%%%%%%%
Following Ref.~\cite{Rubakov:1983bb}, we introduce the Lagrangian for a bulk fermion $\psi$ coupled to
  the scalar field $\varphi$, 
\bea
\mathcal{L}&=&i \overline{\psi}\left[\gamma^{\mu}D_{\mu}+i\gamma^{5}D_{y}\right]\psi+Y \varphi \overline{\psi}\psi \nonumber \\ 
&=& 
i \overline{\psi_L} \gamma^{\mu}D_{\mu} \psi_{L}+i \overline{\psi_R} \gamma^{\mu}D_{\mu} \psi_{R}  
- \overline{\psi_L} D_{y}\psi_{R}+\bar{\psi}_{R}D_{y} \psi_L + Y \varphi \left( \overline{\psi_L}\psi_{R}+ \overline{\psi_R}\psi_{L}\right), 
\label{DM_fermion}
\eea
where $D_M$ is the covariant derivative under the gauge group, $Y$ is a positive constant, and in the second line, 
  we have decomposed the Dirac fermion $\psi$ into its chiral components, 
  $\psi= P_L \psi + P_R \psi \equiv  \psi_L + \psi_R$. 
Neglecting the gauge interaction and replacing $\varphi$ by the kink solution $\varphi_{\rm kink}$, 
  the equations of motion are given by 
\bea
&& i\gamma^{\mu} \partial_{\mu}\psi_{L}- \partial_{y}\psi_{R} + Y \varphi_{\rm kink} \psi_{R}=0,  \nonumber \\
&& i\gamma^{\mu}  \partial_{\mu}\psi_{R}+ \partial_{y}\psi_{L} + Y \varphi_{\rm kink} \psi_{L}=0.
\label{EOM_F1}
\eea
Using the KK-mode decompositions, 
\bea
 \psi_L (x,y) = \sum_{n=0}^\infty \psi_L^{(n)}(x) \, \chi_L^{(n)}(y), \; \; \;
 \psi_R (x,y) =\sum_{n=0}^\infty \psi_R^{(n)}(x) \, \chi_R^{(n)}(y), 
\eea
  we obtain the following KK-mode equations: 
\bea
&& \left[ -\partial_y^2 - \frac{\gamma_F (\gamma_F+1) m_\varphi^2}{\cosh^2(m_\varphi y)} \right] \chi_L^{(n)}
  = \left(M_n^2 - \gamma_F^2 m_\varphi^2 \right)  \chi_L^{(n)}, \nonumber\\
&& \left[ -\partial_y^2 - \frac{\gamma_F (\gamma_F-1) m_\varphi^2}{\cosh^2(m_\varphi y)} \right] \chi_R^{(n)}
  = \left(\overline{M}_n^2 - \gamma_F^2 m_\varphi^2 \right)  \chi_R^{(n)},   
\label{KK_EOM4}
\eea
where $\gamma_F=Y/\sqrt{\lambda}$. 
Again, these equations are equivalent to the time-independent Schr\"odinger equation 
  in one-dimensional quantum mechanics.

We are interested in localized fermions, in other words, bound states 
  satisfying the boundary conditions, $\chi_{L, R}^{(n)} \to 0$ for $y \to 0$.  
Such solutions can be expressed in terms of hypergeometric functions $F[a,b; c; y]$ 
  (see, e.g., Ref.~\cite{Hisano:1999bn}). 
The mass eigenvalues for $\chi_{L}^{(n)}$ are given by 
\bea 
  M_n^2 = n \left( 2 \gamma_F -n \right) m_\varphi^2  \; \; \; \; (n=0, 1, 2, \cdots < \gamma_F).
\label{mn1}  
\eea 
Since bound state eigenvalues must satisfy $E_n = m_n^2 - \gamma_F^2 m_\varphi^2 < 0$, 
  the number of localized KK-modes is finite, and a massive mode exists only for $\gamma_F > 1$. 
The eigenfunctions for even values of 
  $n= 2 n^\prime \; (n^\prime =0,1,2, \ldots )$ and odd values of 
  $n= 2 n^{\prime \prime}+1 \; (n^{\prime \prime} =0,1,2,\ldots )$ 
  are given by (up to normalization factors) 
\begin{eqnarray} 
\chi_{L}^{(n^\prime)}(y) = \left[ \cosh(m_\varphi y) \right]^{- \gamma_F} \, 
   F\left[-n^\prime , -\gamma_F + n^\prime; 1/2 ; 1- \cosh^2(m_\varphi y) \right], 
\end{eqnarray}
and 
\begin{eqnarray}
  \chi_{L}^{(n^{\prime \prime})}(y) =  \sinh(m_\varphi y) 
      \left[\cosh(m_\varphi y) \right]^{- \gamma_F}  \, 
  F\left[-n^{\prime \prime}, - \gamma_F + n^{\prime \prime}+1 ; 3/2 ; 1- \cosh^2(m_\varphi y) \right],   
\end{eqnarray}
respectively. 
Similarly, the mass eigenvalues for $\chi_{R}^{(n)}$ are given by 
\bea 
  \overline{M}_n^2 = (n+1) \left( 2 \gamma_F -(n+1) \right) m_\varphi^2  \; \; \; \; (n=0, 1, 2, \cdots < \gamma_F-1).
\eea 
The eigenfunctions for even valuess of 
  $n= 2 n^\prime \; (n^\prime =0,1,2, \ldots )$ and odd values of 
  $n= 2 n^{\prime \prime}+1 \; (n^{\prime \prime} =0,1,2,\ldots )$ 
are given by (up to normalization factor) 
\begin{eqnarray} 
\chi_{R}^{(n^\prime)}(y)= \left[ \cosh(m_\varphi y) \right]^{- \gamma_F+1} \, 
   F\left[-n^\prime , -\gamma_F + n^\prime+1; 1/2 ; 1- \cosh^2(m_\varphi y) \right], 
\end{eqnarray}
and 
\begin{eqnarray}
\chi_{R}^{(n^{\prime \prime})}(y) =  \sinh(m_V y) 
      \left[\cosh(m_\varphi y) \right]^{- \gamma_F+1}  
  F\left[-n^{\prime \prime}, - \gamma_F + n^{\prime \prime}+2 ; 3/2 ; 1- \cosh^2(m_\varphi y) \right],   
\end{eqnarray}
respectively. 
Note that the mass eigenvalue $\overline{M_n}$ for $\chi_R^{(n)}$ coincide with the $(n+1)$-th mass eigenvalue for $\chi_L^{(n)}$, 
  namely, $\overline{M}_n=M_{n+1}$. 
This implies that only one (left-handed) chiral fermion emerges in the four-dimensional effective theory,
   which can be identified as an SM fermion when we extend this setup to the five-dimensional Standard Model.    
The KK-modes of $\chi_L^{(n)}$ and $\chi_R^{(n-1)}$ for $n \geq 1$ pair up to form Dirac fermions in the four-dimensional theory.
For convenience, we relabel $\chi_R^{(n-1)}$ as ``$\chi_R^{(n)}$" to align with its Dirac mass partner $\chi_L^{(n)}$,
   and we will use this notation hereafter.
It is straightforward to see that changing $Y \to -Y$ in Eq.~(\ref{EOM_F1}) leads to 
   Eq.~(\ref{KK_EOM4}) with the replacement $\chi_L^{(n)} \leftrightarrow \chi_R^{(n)}$,
   resulting in a right-handed massless fermion instead of a left-handed one.

For simplicity in the phenomenological analysis in the next section, we set $\gamma_F=Y/\sqrt{\lambda} =2$. 
In this case, only one KK-mode Dirac fermion appears in the four-dimensional effective theory,
  with mass $m_{\rm KK} = \sqrt{3} m_\varphi$.
The explicit forms of the KK-mode functions are given by
\bea
 \psi_L (x,y) &=& 
  \frac{\sqrt{3 m_\varphi} }{2} \left[ \frac{1}{\cosh^2(m_\varphi y)} \right] \psi_L^{(0)}(x)
    + \sqrt{ \frac{3m_\varphi}{2} } \left[ \frac{\sinh (m_\varphi y)}{\cosh^2 (m_\varphi y)} \right] \psi_L^{(1)}(x),   \nonumber \\
 \psi_R (x,y) &=&  \sqrt{ \frac{m_\varphi}{2}} \left[ \frac{1}{\cosh(m_\varphi y)} \right] \psi_R^{(1)}(x), 
\label{F_example} 
\eea
with the kinetic terms canonically normalized in the four-dimensional effective theory. 
Interestingly, in this choice of $\gamma_F=2$, the KK-mode wavefunctions for $\psi_L$ coincide 
  with those of the scalar field $\varphi$ in Eq.~(\ref{s-mode}).

%%%%%%%%%%%%%%%%%%%%%%%%%%%%%%
\subsection{Domain-Wall SM Fermions and NG-mode Couplings}
\label{sec:2-C}
%%%%%%%%%%%%%%%%%%%%%%%%%%%%%%
%%%%%%%%%%%%%%%%%%%%%%%%%%%%%
\begin{table}[h]
    \centering
    \begin{tabular}{|c|c|c|}
    \hline
    & SU(2)$_L$ & U(1)$_{Y}$ \\
    \hline
    $\Psi_{D}$ & $\mathbf{2}$ & $-\frac{1}{2}$ \\
    $\Psi_{S}$ & $\mathbf{1}$ & $-1$ \\
    \hline
    $H$ & $\mathbf{2}$ & $\frac{1}{2}$ \\
    \hline
    $\varphi$ & $\mathbf{1}$ & $0$ \\
    \hline
    \end{tabular}
    \caption{Particle content for the first-generation lepton sector in the five-dimensional bulk. 
    }
    \label{tab:1}
\end{table}
%%%%%%%%%%%%%%%%%%%%%%%%%%%%%%

We now extend the domain-wall fermion framework to the SM fermions.
For simplicity, we focus on the first-generation lepton sector in the five-dimensional bulk. 
The particle content is summarized in Table~\ref{tab:1}.
In addition to the lepton doublet, lepton singlet, and Higgs doublet, 
   we also include the kink scalar $\varphi$ discussed in Section~\ref{sec:2-A}.

The relevant part of Lagrangian is given by 
\bea
\mathcal{L}_5 &= & 
       i \, \overline{\Psi_D} \left( \gamma^\mu D_\mu^D + i \gamma^5 D_y^{D} \right) \Psi_{D} 
       + Y_D \, \varphi \, \overline{\Psi_D} \Psi_D  \nonumber\\
&+&  i \, \overline{\Psi_S} \left( \gamma^\mu D_\mu^S + i \gamma^{5} D_y^S \right) \Psi_S 
       - Y_S \, \varphi \, \overline{\Psi_S} \Psi_S  \nonumber \\
&-&  Y_e \, \overline{\Psi_D} H \Psi_S + \text{h.c.} , 
\eea
where $D_M^{D, S}$ represent the covariant derivatives for the electroweak gauge group, 
  and all Yukawa coupling constants are taken to be real and positive.  
Decomposing the fermions into their chiral components, $\Psi = \Psi_{L} + \Psi_{R}$,
  the Lagrangian becomes
\bea
  \mathcal{L}_5 &=& i \; \overline{\Psi_{DL}} \gamma^\mu D_\mu^D \Psi_{DL} 
   + i \; \overline{\Psi_{DR}} \gamma^\mu D_\mu^D \Psi_{DR} - \bar{\Psi}_{DL} D_y \Psi_{DR} +\bar{\Psi}_{DR} D_y \Psi_{DL} \nonumber\\
    &+& Y_{D} \varphi \left( \; \overline{\Psi_{DL}} \Psi_{DR} + \; \overline{\Psi_{DR}} \Psi_{DL} \right) \nonumber\\
    &+& \; i \; \overline{\Psi_{SL}} \gamma^\mu D_\mu^D \Psi_{SL} + i \; \overline{\Psi_{SR}} \gamma^\mu D_\mu^D \Psi_{SR}
     - \overline{\Psi_{SL}} D^{D}_y \Psi_{SR} + \overline{\Psi_{SR}} D^{D}_y \Psi_{SL} \nonumber\\
    & - & \; Y_S \varphi \left( \overline{\Psi_{SL}} \Psi_{SR} + \overline{\Psi_{SR}} \Psi_{SL} \right) \nonumber\\
    & - & \left( Y_e\; \overline{\Psi_{DL}} H \Psi_{SR} - Y_e \; \overline{\Psi_{DR}} H \Psi_{SL} +{\rm h.c.} \right). 
\eea
For simplicity, we set $\gamma_F= Y_D/\sqrt{\lambda}=Y_S/\sqrt{\lambda}=2$
   so that the KK-mode decomposition follows Eq.~(\ref{F_example})). 
For the lepton doublet, 
\bea
\Psi_{DL}(x, y) &=& \frac{\sqrt{3} m_y}{2} \frac{1}{\mathrm{cosh}^{2}(m_\varphi y)} \Psi^{(0)}_{DL}(x) + \sqrt{\frac{3 m_y}{2}}    
    \frac{\mathrm{sinh}(m_\varphi y)}{\mathrm{cosh}^2(m_\varphi y)} \Psi^{(1)}_{DL}(x) , \nonumber\\
\Psi_{DR}(x, y) &=& \sqrt\frac{{m_y}}{2} \frac{1}{\mathrm{cosh}(m_\varphi y)} \Psi^{(1)}_{DR}(x), 
    \label{eq:KK-D}
\eea
while for the lepton singlet, 
\bea
 \Psi_{SL}(x, y) &=& \sqrt\frac{{m_y}}{2} \frac{1}{\mathrm{cosh}(m_\varphi y)} \Psi^{(1)}_{SL}(x) ,  \nonumber\\ 
 \Psi_{SR}(x, y) &=& \frac{\sqrt{3} m_y}{2} \frac{1}{\mathrm{cosh}^{2}(m_\varphi y)} \Psi^{(0)}_{SR}(x) + \sqrt{\frac{3 m_y}{2}} \frac{\mathrm{sinh}(m_\varphi y)}{\mathrm{cosh}^2(m_\varphi y)} \Psi^{(1)}_{SR}(x). 
    \label{eq:KK-S}
\eea
Here, $\Psi^{(0)}_{DL}$ and $\Psi^{(0)}_{SR}$ are identified with the SM lepton doublet and singlet, respectively, 
  and the first KK-mode fermion mass is $m_{\rm KK} = \sqrt{3} m_{\varphi}$.

By substituting these KK-mode functions, the kink scalar mode from Eq.~(\ref{s-mode}),
   and the Higgs vacuum expectation value (VEV) $\langle H \rangle = (0 \; v/\sqrt{2})^T$ into $\mathcal{L}_5$,
   and then integrating over $y$, we obtain the four-dimensional effective Lagrangian.
The kinetic terms in the effective Lagrangian are canonically normalized: 
\bea
    \mathcal{L}^{kin}_{4D} &=& 
    \; i \; \overline{{\Psi}^{(0)}_{DL}} \gamma^\mu D_\mu^D \Psi^{(0)}_{DL} 
   + i \; \overline{{\Psi}^{(1)}_{DL}} \gamma^\mu D_\mu^D \Psi^{(1)}_{DL} 
   + i \; \overline{{\Psi}^{(1)}_{DR}} \gamma^\mu D_\mu^D \Psi^{(1)}_{DR}  \nonumber\\
 &+& \; i \; \overline{{\Psi}^{(0)}_{SR}} \gamma^\mu D_\mu^D \Psi^{(0)}_{SR} 
         + i \; \overline{{\Psi}^{(1)}_{SL}} \gamma^\mu D_\mu^D \Psi^{(1)}_{SL} 
         + i \; \overline{{\Psi}^{(1)}_{SR}} \gamma^\mu D_\mu^D \Psi^{(1)}_{SR} .
\eea
The fermion mass terms are found to be
\bea
  \mathcal{L}^{Mass}_{4D} & =& 
         m_{\rm KK} \left( \overline{{\Psi}^{(1)}_{DL}} \Psi^{(1)}_{DR} + {\rm h.c.} \right) 
      -  m_{\rm KK} \left( \overline{{\Psi}^{(1)}_{SL}} \Psi^{(1)}_{SR} + {\rm h.c.} \right) \nonumber \\
 &-& Y_e \left( \overline{{\Psi}^{(0)}_{DL}} \langle H \rangle \Psi^{(0)}_{SR} +{\rm h.c.} \right) 
  -Y_e  \left( \overline{{\Psi}^{(1)}_{DL}}  \langle H \rangle \Psi^{(1)}_{SR} +  \bar{\Psi}^{(1)}_{SL} 
    \langle H \rangle^\dagger \Psi^{(1)}_{DR} + {\rm h.c.} \right). 
\eea
Note that there are no mixing mass terms between the SM leptons and their first KK-mode fermions. 
Defining $\Psi_D = \begin{bmatrix} \nu_D \ e_D \end{bmatrix}^T$ and $\Psi_S = e_S$,
  the mass terms can be rewritten as
\bea
    \mathcal{L}^{Mass}_{4D} &=&
     m_{\rm KK} \left( \overline{{\nu}^{(1)}_{DL}} \nu^{(1)}_{DR}  +  \overline{{e}^{(1)}_{DL}} e^{(1)}_{DR}  + {\rm h.c.} \right) 
     -  m_{\rm KK} \left(\overline{{e}^{(1)}_{SL}} e^{(1)}_{SR}  + {\rm h.c.} \right) \nonumber\\
     &-& m_{f} \left( \overline{{e}^{(0)}_{DL} e^{(0)}_{SR}}  + {\rm h.c.} \right)  
      - m_{f} \left(  \overline{{e}^{(1)}_{DL}} e^{(1)}_{SR}  + \; \overline{{e}^{(1)}_{SL}} e^{(1)}_{DR}  + {\rm h.c.} \right), 
\label{KK_masses}
\eea
where $m_{f}=\frac{Y_e v}{\sqrt{2}}$. 
The effective Lagrangian also includes Yukawa couplings with the NG boson and the first KK-mode of the kink scalar:
\bea
    \mathcal{L}^{Y}_{4D} &=& 
     y_{\varphi} \left( \overline{{\Psi}^{(0)}_{DL}} \Psi^{(1)}_{DR} + {\rm h.c.} \right)\varphi^{(0)}   
     +   y^{\prime}_{\varphi} \left( \overline{{\Psi}^{(1)}_{DL}} \Psi^{(0)}_{DR} + {\rm h.c.} \right)\varphi^{(1)} \nonumber\\ 
   &-& y_{\varphi} \left(  \overline{{\Psi}^{(0)}_{SR}} \Psi^{(1)}_{SL} + {\rm h.c.} \right)\varphi^{(0)} 
    -  y^{\prime}_{\varphi} \left( \overline{{\Psi}^{(1)}_{SR}} \Psi^{(1)}_{SL} + {\rm h.c.} \right)\varphi^{(1)} \nonumber \\
  &\supset&   
  y_{\varphi} \left( \overline{{\nu}^{(0)}_{DL}} \nu^{(1)}_{DR}  +  \overline{{e}^{(0)}_{DL}} e^{(1)}_{DR}  + {\rm h.c.} \right) \varphi^{(0)} 
  -  y_{\varphi} \left(\overline{{e}^{(0)}_{SR}} e^{(1)}_{SL}  + {\rm h.c.} \right) \varphi^{(0)} ,
\label{NG_Yukawa}
\eea
where $y_{\varphi}=\frac{9\pi\sqrt{2m_\phi} Y_{D/S}}{64}$ 
  and $y^{\prime}_{\varphi}=\frac{3\pi\sqrt{2m_\phi} Y_{D/S}}{32}$. 
In this paper, we focus on phenomenology driven by the Yukawa couplings of NG boson $\varphi^{(0)}$ 
 with SM fermions and their KK-modes.

For the analysis in the next section, we rewrite the Yukawa interactions in terms of KK-mode mass eigenstates. 
The electron KK-mode mass matrix in Eq.~(\ref{KK_masses}),
\bea
 \mathcal{L}^{Mass}_{4D} & \supset& 
 - \begin{pmatrix}
     \; \overline{e^{(1)}_{DL}} &  \; \overline{{e}^{(1)}_{SL} }
  \end{pmatrix}
  \begin{pmatrix}
    -m_{\rm KK} & m_{f} \\
    m_{f} & m_{\rm KK}
  \end{pmatrix}
  \begin{pmatrix}
    e^{(1)}_{DR} \\
    e^{(1)}_{SR}
  \end{pmatrix}
  + {\rm h.c.}
\eea
can be diagonalized by defining mass eigenstates: 
\bea
  \begin{pmatrix}
    e_{D{L}}^{(1)} \\
    e_{S{L}}^{(1)}
  \end{pmatrix}
  =
  \begin{pmatrix}
    \cos \alpha & \sin \alpha \\
    -\sin \alpha & \cos \alpha
  \end{pmatrix}
  \begin{pmatrix}
    E_{1{L}} \\
    E_{2{L}}
  \end{pmatrix} , \; \;
\begin{pmatrix}
    e_{D{R}}^{(1)} \\
    e_{S{R}}^{(1)}
  \end{pmatrix}
  =
  \begin{pmatrix}
    \cos \alpha & \sin \alpha \\
    - \sin \alpha & \cos \alpha
  \end{pmatrix}
  \begin{pmatrix}
    -1 & 0 \\
    0 & +1
  \end{pmatrix}
  \begin{pmatrix}
    E_{1{R}} \\
    E_{2{R}} 
  \end{pmatrix}, 
\eea
where the mixing angle $\alpha$ satisfies  $\tan(2 \alpha) = m_f / m_{\rm KK}$.    
The eigenstates $E_1$ and $E_2$ have the same mass of $M=\sqrt{m_{\rm KK}^{2}+m_{f}^{2}}$. 
In terms of the mass eigenstates, we rewrite Eq.~(\ref{NG_Yukawa}) as
\bea
\mathcal{L}^{Y}_{4D} &\supset&
 -y_{\varphi} \, \overline{e}\, (V_{1}+A_{1}\gamma_{5})\, E_{1}\, \varphi^{0} 
 - y_{\varphi}\, \overline{{E}_{1}} \, (V_{1} -A_{1}\gamma_{5}) \, e  \, \varphi^{0} 
  \nonumber\\ 
&&- y_{\varphi}\, \overline{e}\, (V_{2}+A_{2}\gamma_{5}) \, E_{2} \, \varphi^{0}  
- y_{\varphi}\, \overline{{E}_{2}} \, (V_{2}- A_{2}\gamma_{5}) \, e \, \varphi^{0}, 
\label{NG_Yukawa-2}
\eea
where we defined $e^{(0)}_{DL} = P_L e$ and $e^{(0)}_{SR} = P_R e$ with $e$ being the SM electron, and
\bea
 V_1 = V_2 = \frac{1}{2} (\cos \alpha - \sin \alpha), \; \;   A_{1} = - A_{2} = \frac{1}{2} (\cos \alpha + \sin \alpha). 
\eea

Although we explicitly discussed only the lepton sector of the Domain-Wall SM,
  the corresponding formulas for the quark sector can be easily derived in the same manner.

%%%%%%%%%%%%%%%%%%%%%%%%%%%%%%%%%%%%%%%%%%%%%%%%%%%%%%%%%%
\section{Phenomenological constraints on the NG-mode physics}
\label{sec:3}
%%%%%%%%%%%%%%%%%%%%%%%%%%%%%%%%%%%%%%%%%%%%%%%%%%%%%%%%%%
Using the formulas presented in the previous section, we consider the phenomenology involving the NG boson
   associated with the spontaneous breaking of the translational invariance in the fifth-dimensional direction. 
Through the Yukawa couplings in Eq.~(\ref{NG_Yukawa-2}), we may consider the processes of the SM fermion pair annihilation 
   to a pair of the NG bosons: $f_{SM} \overline{f_{SM}} \to \varphi^{(0)} \varphi^{(0)}$. 
Such processes can occur in various environments, such as the cores of supernovae and the early universe. 

To calculate the cross section of the process $e^+ e^- \to \varphi^{(0)} \varphi^{(0)}$, we consider $t$ and $u$-channel processes 
  mediated by $E_1$ and $E_2$ with the degenerate mass $M = \sqrt{m_{\rm KK}^2 + m_e^2}$. 
The invariant amplitude squared is given by 
\bea
\sum_{\rm spins} |{\mathcal M}_t+{\mathcal M}_u|^2 
&=&  \frac{1}{2}  \frac{ y_\varphi^4}{(M^2 - t )^2 (M^2-u)^2} (s^2-(t-u)^2)  \nonumber \\
&\times&  \left\{ (M^2-u)^2+ (M^2-t)^2 -2 (M^2-t) (M^2-u ) \right\} \nonumber\\
&=&\frac{1}{2}  \frac{ y_\varphi^4}{(M^2 - t )^2 (M^2-u)^2} (s^2-(t-u)^2)  (t-u)^2, 
\eea
where ${\mathcal M}_t$ and ${\mathcal M}_u$ are the $t$ and $u$-channel amplitudes, respectively, 
  and $s, t, u$ are Mandelstam's variables. 
In the second line, the first, second and third terms are from $|{\mathcal M}_t|^2$, $|{\mathcal M}_u|^2$, 
  ${\mathcal M}_t^\dagger {\mathcal M}_u +{\mathcal M}_u^\dagger {\mathcal M}_t$, respectively. 
We can see that the terms proportional to $M^4$ and $M^2$ from purely $s$ and $t$-channel diagrams 
  are canceled by the interference.  
For $M^2 \gg s, |t|, |u|$, and $m_e^2$, the total cross section is simply given by
\bea
\sigma \simeq \frac{y_\varphi^4}{1920 \pi} \frac{s^3}{m_{\rm KK}^8} , 
\label{Xsec}
\eea  
which is strongly suppressed for $s \ll m_{\rm KK}^2$ due to the cancelation mentioned above.

%%%%%%%%%%%%%%%%%%%%%%%%%%%%
\subsection{Constraint from supernova 1987A}
%%%%%%%%%%%%%%%%%%%%%%%%%%%%
If the massless NG boson produced from $e^+ e^-$ annihilations in the core of a supernova (SN) escape from the core, 
  it can lead to excessive SN cooling. 
The observations of neutrinos from the SN 1987A by 
  the Kamiokande II and IMB Collaborations \cite{Kamiokande-II:1987idp, Bionta:1987qt}
  indicate that almost all of the gravitational binding energy of the SN is released by the emission of neutrinos. 
Thus, the energy release by the NG bosons should not exceed the one from neutrinos, 
  leading to an upper bound on the emissivity for free-streaming NG bosons. 
For our analysis, we adopt the so-called Raffelt criterion \cite{Raffelt:1996wa} for the upper bound on the emissivity: 
\bea
\frac{d \varepsilon}{d t} <10^{19} \left[ \mathrm{erg} / \mathrm{g} / \mathrm{s}\right]\times \rho_{SN}\left[ \mathrm{g}/\mathrm{cm}^3 \right]
 \simeq 9.42 \times 10^{-30} \, \left[\mathrm{GeV}^5\right],
\label{Raffelt_Criterion} 
\eea
where $\rho_{SN} \simeq 3 \times 10^{14} \left[\mathrm{g} / \mathrm{cm}^3\right]$ is the SN core density \cite{Burrows:1986me}.

Let us now calculate the emissivity of the NG boson production, the energy emitted by the NG bosons 
  from the SN 1987A per unit time and unit volume, to compare with the upper bound of Eq.~(\ref{Raffelt_Criterion}). 
For the process $e^+ e^- \to \varphi^{(0)} \varphi^{(0)}$, the emissivity is described as
\bea 
\frac{d \varepsilon}{d t}=\int \frac{d^3 p_{e^-}}{(2 \pi)^3} \int \frac{d^3 p_{e^+}}{(2 \pi)^3} \,
 f_{e^-} f_{e^+} \,  \sigma v_{\text {rel }}\left(E_{e^-} + E_{e^+}\right),
 \label{Emissivity1}
\eea
where $v_{\text {rel}}$ is a relative velocity of the colliding $e^\pm$, and 
   the Fermi-Dirac distributions for $e^\pm$ are, respectively, given by
\bea
f_{e^-} &=& \frac{1}{e^{\frac{E_{e^-} -\mu_e}{T_c}}+1},  \nonumber \\
f_{e^+} &=& \frac{1}{e^{\frac{E_{e^+}+\mu_e}{T_c}}+1}, %\simeq \frac{e^{-\mu_e / T_c}}{e^{E_{e^+} / T_c}}, 
\label{FD_distribution}
\eea
with an electron chemical potential $\mu_{e}$. 
In our analysis, we set the number density of electron in the SN core $n_{e^-} \simeq 8.7 \times 10^{37} / \mathrm{cm}^3$ 
  and the core temperature $T_c =30$ MeV $\gg m_e$ \cite{Burrows:1986me}. 
Assuming $\mu_e \gg T_c$ in the SN core, the electrons are in the state of ultra-relativistic degenerate fermion gas, 
  and the the number density of electron is approximately given by
\bea
    n_{e^-} =  2 \int \frac{d^{3} p_{e^-}}{(2 \pi)^3} f_{e^-}  \simeq  \frac{\mu_e^3}{3 \pi^2}.  
\eea
Using $n_{e^-} \simeq 8.7 \times 10^{37} / \mathrm{cm}^3$, we find  $\mu_e \simeq 270$ MeV$ \gg T_c =30$ MeV, 
  which is consistent with the assumption $\mu_e \gg T_c$. 
% 
%We have used $\mu_e \gg T_c$ for $f_{e^+}$ in Eq.~(\ref{FD_distribution}). 
%With this approximation, the positron number density is given by
In this case, the positron number density is given by
\bea
  n_{e^{+}} =   2 \int \frac{d^{3} p_{e^+}}{(2 \pi)^3} f_{e^+}  
  \simeq 2 e^{-\mu_e / T_c} \int \frac{d^{3} p_{e^+}}{(2 \pi)^3} e^{-p_{e^+} / T_c}=\frac{2}{\pi} T_c^3 e^{-\mu_e / T_c}, 
\eea
   which is highly suppressed compared to $n_{e^-}$. 
      
To proceed with the evaluation of Eq.~(\ref{Emissivity1}), we define  
    $E_{\pm} =E_{e^-} \pm E_{e^+}$  along with their kinematic ranges,  $E_{+} \geq \sqrt{s}$ and 
     $\left|E_{-}\right| \leq \sqrt{E_{+}^2-s}$, to obtain
\bea 
\frac{d \varepsilon}{d t}=\frac{\pi^2}{(2 \pi)^6}  \int d s \, \sigma(s) s \int d E_{+} \, d E_{-} \, f_{e^-} \,  f_{e^+} \,  E_{+}, 
\label{Emissivity2}
\eea
where we have used 
\bea
 \int d^3 p_{e^-} \, d^3 p_{e^+} \, \sigma v_{\text {rel }} 
=  2 \pi^2 \int  d s \,  d E_{+} \, d E_{-} \, E_{e^-} \, E_{e^+} \, \sigma(s) v_{\text {rel }} 
=  \pi^2 \int  d s \,  d E_{+} \, d E_{-}  \, \sigma(s) \, s, 
\eea     
by neglecting $m_e$. 
Using $f_{e^-}  f_{e^+} \simeq 
  \left( e^{E_+ / T_c}+e^{\mu_e / T_c} e^{\frac{E_+-E_-}{2 T_c}} \right)^{-1}$ from Eq.~(\ref{FD_distribution}) 
  and Eq.~(\ref{Xsec}) for the explicit form of the cross section, 
  we can numerically integrate Eq.~(\ref{Emissivity2}) to arrive at
\bea  
  \frac{d \varepsilon}{d t} \simeq    4.85 \times 10^{-19} \left[{\rm GeV}^5\right] \times  \left(\frac{T_c}{30 \, {\rm MeV}} \right)^{13} \, y_\varphi^4 
   \, \left( \frac{1 \, {\rm GeV}}{m_{\rm KK}} \right)^8 .  
\eea
Applying the Raffelt criterion with $T_c=30$ MeV, we find the SN 1987A constraint to be
\bea
  y_\varphi < 2.10 \times 10^{-3} \, 
   \left( \frac{m_{\rm KK}}{1 \, {\rm GeV}} \right)^2. 
\label{Emissivity_Cond}   
\eea

In deriving the condition of Eq.~(\ref{Emissivity_Cond}), we implicitly assumed 
  that all produced NG bosons escape from the SN core. 
If NG bosons are trapped inside the core, they do not contribute to the extra cooling effect on the SN 1987A, 
  and hence, the Raffelt criterion can not be applied. 
To judge whether the produced NG boson are trapped or not, we calculate the mean free path ($\lambda$) of the NG boson.
Since the typical core radius of supernovae is ${\cal O}(10)$ km \cite{Burrows:1986me}, 
  we impose $\lambda > 10$ km for the produced NG bosons to cause the extra cooling effect.  

The produced NG bosons can scatter with $e^\pm$ in the SN core. 
Since $n_{e^-} \gg n_{e^+}$, we focus on the process $e^{-}\varphi^{(0)} \rightarrow e^{-}\varphi^{(0)}$. 
Neglecting $m_e$, the cross section of this scattering process is found to be
\bea
\sigma (e^{-}\varphi^{(0)} \rightarrow e^{-}\varphi^{(0)} ) \simeq \frac{17 \, y_\varphi^4 \, s^3}{192 \pi \, m_{\rm KK}^8}, 
\eea
which is similar to Eq.~(\ref{Xsec}) as expected. 
Since $e^{-}$ is highly degenerate with $\mu_e \gg T_c$, the electron states with $E_{e^{-}}<\mu_e$ are all filled, 
  so that we should consider the Pauli blocking effect in our calculation. 
The mean free path is evaluated by 
\bea
\lambda^{-1}=\int \frac{d^3 p_{e^-}}{(2 \pi)^3} \, f_{e^-} \, \sigma (e^{-}\varphi^{(0)} \rightarrow e^{-}\varphi^{(0)} ) \times \left(1-f_{e^-} \right),
\eea
where the Pauli blocking term $1-f_{e^-}$ suppresses the production of $e^-$ population with $E_{e^{-}}<\mu_e$. 
Using $d^3 p_{e^-} =2 \pi \, E_{e^-}^2 \, d E_{e^-} \, d (\cos \theta)$ and  $s=2 E_{e^-} \, E_{\varphi^{(0)}} (1-\cos \theta)$ 
  with the NG boson energy $E_{\varphi^{(0)}}$ and the scattering angle $\theta$, 
  we can numerically perform the phase space integral and obtain  
\bea
 \lambda^{-1} \simeq 1.42 \times 10^{-6} \left[ {\rm GeV} \right] \times y_\varphi^4  
 \left( \frac{T_c}{30 \, {\rm MeV}} \right)^6  
 \left( \frac{E_{\varphi^{(0)}}}{1 \, {\rm GeV}} \right)^3
 \left( \frac{1 \, {\rm GeV}}{m_{\rm KK}} \right)^8 .
\eea
Therefore, the condition of $\lambda > 10$ km leads to 
\bea
   y_\varphi< 3.44 \times 10^{-4} \left( \frac{m_{\rm KK}}{1 \, {\rm GeV}} \right)^2
    \left( \frac{1 \, {\rm GeV}}{E_{\varphi^{(0)}}} \right)^{3/4}
    \simeq 9.17 \times 10^{-4} \, \left( \frac{m_{\rm KK}}{1 \, {\rm GeV}} \right)^2, 
\label{Trapping}    
\eea
where we have set $T_c=30$ MeV and $E_{\varphi^{(0)}}=\mu_e$ 
  since the NG bosons are produced via $e^+e^-\to \varphi^{(0)}\varphi^{(0)}$ 
  with the highly degenerate electrons with $E_{e^-} \lesssim \mu_e$.

%%%%%%%%%%%%%%%%%%%%%%%%%%%%
\subsection{BBN bound}
%%%%%%%%%%%%%%%%%%%%%%%%%%%%
In the early universe, the NG boson remains in thermal equilibrium with the SM particle plasma 
   through the process $f_{SM} \overline{f_{SM}} \leftrightarrow \varphi^{(0)} \varphi^{(0)}$. 
As the universe evolves, the NG boson eventually decouples from the SM thermal bath at a certain temperature.
To preserve the success of the BBN for the synthesis of light nuclei, 
   the energy density of the NG boson must remain below a certain upper limit at the BBN era 
   characterized by a temperature of the SM plasma $T_{\rm BBN} \simeq 1$ MeV.

Suppose that the NG boson decouples from the SM plasma at $T_{\rm dec} > T_{\rm BBN}$. 
Considering the entropy conservation of the SM plasma and the decoupled NG boson system
   during the adiabatic evolution of the universe,  
   the temperature of the NG boson system at the BBN era ($T_{\rm BBN}^\prime $) is evaluated as 
\bea
  T_{\rm BBN}^\prime =\left(  
    \frac{g_*^{SM} (T_{\rm BBN})}{g_*^{SM} (T_{\text {\rm dec }})} 
    \right)^{1/3} T_{\rm BBN} , 
\label{Tdef}
\eea   
where $g_*^{SM}(T)$ is the total relativistic degrees of freedom of the SM plasma at temperature $T$. 
The extra energy density contribution from the NG boson system at the BBN era is parameterized 
   in terms of the number of extra neutrino species $\Delta N_{\rm eff}$ by 
\bea
\rho_{\varphi^{(0)}}=\frac{\pi^2}{30}\left(T_{\rm BBN}^{\prime}\right)^4 
 = \frac{\pi^2}{30} \, \frac{7}{4} \, \Delta N_{\text {eff }}\left(T_{\rm BBN}\right)^4 
 \label{BBN}
\eea  
Using the Planck 2018 result \cite{Planck:2018vyg},
\begin{eqnarray}
    N_{\rm eff} = 2.99 \pm 0.17
\end{eqnarray}
    at 68\% C.L., and the SM prediction $N_{\rm eff}^{\rm SM} = 3.046$ 
   (see, for example, Refs.~\cite{Escudero:2018mvt,Bennett:2019ewm,Escudero:2020dfa,Akita:2020szl,Bennett:2020zkv}),     
   we obtain the upper bound on $\Delta N_{\rm eff} \leq 0.114$.     
From Eqs.~(\ref{Tdef}) and (\ref{BBN}), together with $g_*^{SM}(T_{\rm BBN})=10.75$, 
  this upper bound is interpreted to a lower bound on $g_*^{SM} (T_{\rm dec}) \geq 36$, 
  which corresponds to a lower bound on  $T_{\rm dec} \gtrsim 200$ MeV (around the QCD phase transition era).

We now consider the decoupling of the NG boson from the SM thermal plasma. 
In deriving a constraint on $y_\varphi$ and $m_{\rm KK}$,  the BBN constraint on $T_{\rm dec} \geq 0.2$ GeV 
  is expressed as 
\bea  
   n_{\varphi^{(0)}}^{\rm eq}  \langle \sigma (\varphi^{(0)} \varphi^{(0)} \to f_{SM} \overline{f_{SM}}) v_{\rm rel}\rangle  \leq H
\label{BBN_Condition}   
\eea  
at $T=0.2$ GeV, where $n_{\varphi^{(0)}}^{\rm eq}=\frac{\zeta(3)}{\pi^2} T^3$ is the thermal equilibrium number density of the NG boson, 
   $\langle \sigma (\varphi^{(0)} \varphi^{(0)} \to f_{SM} \overline{f_{SM}}) v_{\rm rel}\rangle$ is the thermally averaged 
   annihilation cross section of the NG boson pair times relative velocity $v_{\rm rel}$, 
   and $H=\sqrt{\frac{\pi^2}{90} (g_*^{SM}(T)+1)} \frac{T^2}{M_P}$ is the Hubble parameter 
   with the reduced Planck mass of $M_P=2.43 \times 10^{18}$ GeV. 
In calculating the NG boson annihilation cross section at $T=0.2$ GeV, 
   we consider the following final states: a pair of neutrino ($\nu_{\alpha}$, $\alpha=e, \mu, \tau$), 
    $\ell^+ \ell^-$ ($\ell=e, \mu$) and $q \bar{q}$ ($q=u, d$). 
Neglecting the SM fermion masses for simplicity, we find 
\bea 
&&\sum_f \sigma\left(\varphi^{(0)} \varphi^{(0)} \rightarrow f \bar{f} \right)   \nonumber\\
&=& \sum_{\alpha=e, \mu,  \tau} \sigma\left(\varphi^{(0)} \varphi^{(0)} \rightarrow \nu_\alpha \bar{\nu}_\alpha \right) 
+
\sum_{\ell=e, \mu} \sigma\left(\varphi^{(0)} \varphi^{(0)} \rightarrow \ell^+ \ell^- \right)  
+
\sum_{q=u, d} \sigma\left(\varphi^{(0)} \varphi^{(0)} \rightarrow q \bar{q} \right)  \nonumber\\
&=& (1 \times 3 + 2 \times 2 + 6 \times 2 ) \sigma\left(\varphi^{(0)} \varphi^{(0)} \rightarrow \nu_\alpha \bar{\nu}_\alpha \right) 
= 19 \times \frac{y_\varphi^4}{960 \pi} \frac{s^3}{ m_{\rm KK}^8}, 
%
%=\frac{y_\varphi^4}{960 \pi} \frac{s^3}{ m_{\rm KK}^8}, 
%\nonumber \\
%
%&&\sigma\left(\varphi^{(0)} \varphi^{(0)} \rightarrow \nu_\alpha \bar{\nu}_\alpha \right)  =\frac{y_\varphi^4}{960 \pi} \frac{s^3}{ m_{\rm KK}^8}, 
%\nonumber \\
%&&\sigma\left(\varphi^{(0)} \varphi^{(0)} \rightarrow \ell^+ \ell^- \right)  = 
%  2 \times \sigma\left(\varphi^{(0)} \varphi^{(0)} \rightarrow \nu_\alpha \bar{\nu}_\alpha \right), 
%\nonumber \\  
%&&\sigma\left(\varphi^{(0)} \varphi^{(0)} \rightarrow q \bar{q} \right)  = 
%  6 \times \sigma\left(\varphi^{(0)} \varphi^{(0)} \rightarrow \nu_\alpha \bar{\nu}_\alpha \right), 
%\sigma\left(\varphi \varphi \rightarrow l^+ l^{-}\right) & =\frac{y_\varphi^4 s^3}{960 \pi M^8} \times 2 \\
%\sigma(\varphi \varphi \rightarrow q \bar{q}) & =\frac{y_\varphi^4 s^3}{960 \pi M^8} \times 2 \times 3 \\
%\sum_f \sigma(\varphi \varphi \rightarrow q \bar{q}) & =\frac{y_\varphi^4 s^3}{960 \pi M^8}(1 \times 3+2 \times 2+2 \times 3 \times 2) \\
%& =\frac{y_\varphi^4 s^3}{960 \pi M^8} \times 19
\eea
where we have assumed a common mass $m_{\rm KK}$ for intermediate KK neutrinos, KK charged leptons and KK quarks. 
The thermally averaged annihilation cross section is given by 
\bea
\left\langle\sigma v_{\text {vel }}\right\rangle &=& \left(n_{\varphi^{(0)}}^{\rm eq}\right)^{-2} 
\frac{T}{32 \pi^4} \int d s \, \sum_f \sigma\left(\varphi^{(0)} \varphi^{(0)} \rightarrow f \bar{f} \right) \, s^{3 / 2} K_1(\sqrt{s} / T) \nonumber\\
&=&\left(n_{\varphi^{(0)}}^{\rm eq}\right)^{-2} 
\frac{19}{15360 \pi^5} \frac{ y_\varphi^4 \, T^{12}}{m_{\rm KK}^8} \int_0^{\infty} d z \, z^{10} K_1(z), 
\eea
where $K_1$ is the modified Bessel function of the second kind of order one. 
Setting $T=0.2$ GeV and, accordingly, $g_*^{SM}=51.25$, we find that Eq.~(\ref{BBN_Condition}) leads to 
\bea
   y_\varphi  \leq  2.00 \times 10^{-4} \, \left( \frac{m_{\rm KK}}{1 \, {\rm GeV}} \right)^2. 
\label{BBN_bound}   
\eea

%%%%%%%%  Fig 1 %%%%%%%%%%%%%%%%%%%%%%%%%%%%%%%
\begin{figure}[t!]
\begin{center}
\includegraphics[width=0.7\textwidth]{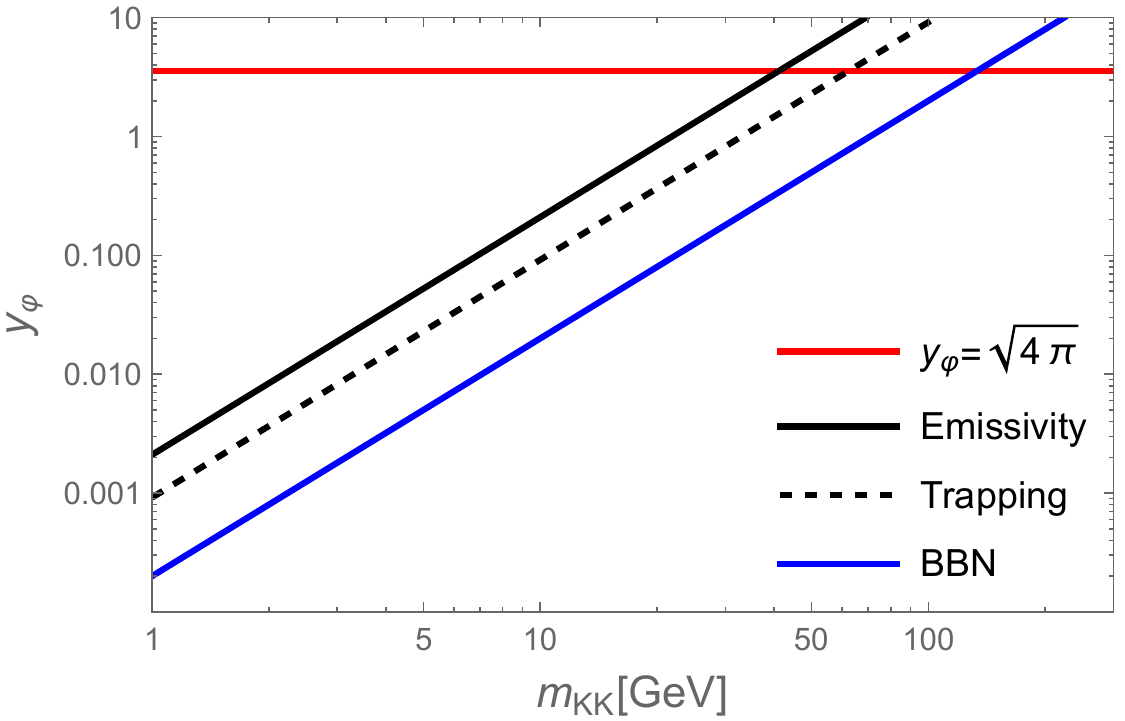} 
\caption{
Summary of constraints from SN 1987A observations, BBN, and coupling perturbativity. 
The black solid line shows the upper limit from SN 1987A cooling (Raffelt criterion), excluding the region above it, 
while the black dashed line indicates the trapping condition limit, with the Raffelt criterion applicable only below this line. 
The blue line is the upper limit to satisfy the BBN bound, while the horizontal red line shows the perturbativity limit $y_\varphi \leq \sqrt{4\pi}$.
} 
\label{fig:1}
\end{center}
\end{figure}
%%%%%%%%%%%%%%%%

In Fig.~\ref{fig:1}, we summarize our results for the SN 1987A constraints and BBN bound,  
  along with a bound from the coupling perturbativity which we have set $y_\varphi \leq \sqrt{4 \pi}$ (horizontal red line). 
The black solid line represents the upper limit on $y_\varphi$ that satisfies the Raffelt criterion in Eq.~(\ref{Emissivity_Cond}) 
  to avoid excessive SN cooling. 
The black dashed line represents the trapping condition in Eq.~(\ref{Trapping}).   
For the parameters on the left side of this line, the produced NG bosons are trapped inside the SN core 
  and the NG bosons do not contribute to an extra cooling effect, 
  while for the parameters on the right side of the line, the NG bosons escapes from the core. 
Note that the dashed line appears on the right side of the solid line. 
This means that for the parameters on the right side of the dashed line, 
  the produced NG bosons contribute to the extra cooling effect but this effect is small enough 
  to be consistent with the SN 1987A observations. 
Since the Raffelt criterion is not applicable for parameters to the left side of the dashed line, 
  entire parameter region below the horizontal red line shown in the figure is consistent with the SN 1987A observations.\footnote{
More precisely, note that for $m_{\rm KK} \lesssim \mu_e$, Eq.~(\ref{Xsec}) is not a good approximation. 
However, we do not consider such a parameter region since as shown in Fig.~\ref{fig:1}, the BBN bound excludes such a parameter region. 
}
The blue solid line represents the Big Bang Nucleosynthesis BBN bound in Eq.~(\ref{BBN_bound}), 
  which establishes the most stringent upper bound on the coupling $y_\varphi$.

%%%%%%%%%%%%%%%%%%%%%%%%%%%% 
\subsection{LHC Bounds on KK-mode Fermion Mass}
%%%%%%%%%%%%%%%%%%%%%%%%%%%%
If kinematically allowed, KK-mode quarks and KK-mode leptons can be pair-produced at high energy colliders, 
  in particular, at the LHC, through their interactions with the SM gauge bosons. 
For example, a pair of KK-mode quark and anti-quark can be produced at the LHC through gluon fusion, 
  while a pair of KK-mode lepton and anti-lepton can be produced through the weak gauge bosons and photon in the $s$-channel. 
Once a KK-mode quark/lepton is produced, it decays to its SM partner quark/lepton and the NG boson. 
Thus, a characteristic signature of KK-mode fermion productions at the LHC is a final state 
  with two SM fermion jets and a missing energy from NG bosons: 
$p p \to f_{KK} \, \overline{f_{KK}}$ followed by $f_{KK} \to f_{SM} + \varphi^{(0)}$ and  $\overline{f_{KK} } \to \overline{f_{SM}} + \varphi^{(0)}$. 
Note that this is analogous to the signature of the sparticles pair productions 
  in the so-called simplified supersymmetric models \cite{Alwall:2008va, Alwall:2008ag, LHCNewPhysicsWorkingGroup:2011mji}, 
  where a pair of squarks/sleptons is produced at the LHC and each of them decays to its SM partner fermion and the lightest neutralino:
$p p \to {\tilde f} \, \overline{\tilde f}$ followed by ${\tilde f} \to f_{SM} + \tilde{\chi}_1^0$ and  $\overline{\tilde f} \to \overline{f_{SM}} + \tilde{\chi}_1^0$. 
The collider physics for these two processes are essentially the same except for the difference in the spins of the involved particles. 
However, note that in our scenario the NG boson is massless, while the neutralino is usually massive. 

In considering the lower mass bounds on KK-mode fermions, we refer the results for squark/slepton search at the LHC. 
For example, the sparticle search results by the ATLAS Collaboration using up to 140/fb luminosity at $\sqrt{s}=13$ TeV 
   is reported in Ref.~\cite{ATLAS:2024lda}. 
From the result for the sbottom pair production, $pp \to {\tilde b} {\tilde b}$ followed by ${\tilde b} \to b + \tilde{\chi}_1^0$, 
   we can read off the lower bound on $m_{\tilde b} \geq1.27$ TeV for $m_{\tilde{\chi}_1^0}=0$. 
Although the production cross section for KK-mode quarks/leptons is typically larger than that for squarks/sleptons by a factor of a few, 
   we adopt this lower bound on sbottom as a conservative lower bound on KK-mode bottom quark mass. 
Similarly, from the result for the lighter stop quark search, we read $m_{\tilde{t}_1} \geq 1.2$ TeV for $m_{\tilde{\chi}_0^1}=0$, 
   which is reinterpreted to the lower bound on KK-mode top quark in our scenario. 
The lower mass bounds on the selectron and smuon are a bit weaker, $m_{\tilde \ell} \geq 700$ GeV ($\ell=e, \mu$) 
   while $m_{\tilde \tau} \geq 500$ GeV for the stau. 
We reinterpret these mass bounds on the sleptons to the upper bound on mass of KK-mode electron, muon and tau in our scenario. 
Comparing the LHC bounds with the results shown in Fig.~\ref{fig:1}, 
   we see that LHC bounds are much more severe than the BBN bound.

If the Yukawa coupling $y_\varphi$ is large, we may consider a pair production of a KK-mode fermion, 
  $f_{SM} \overline{f_{SM}}  \to f_{KK} \overline{f_{KK}}$, mediated by the NG boson in the $t$-channel \footnote{KK mode quark pair productions at the LHC  by NG boson mediation was previously discussed in Ref.~\cite{Arai:2018uoy}. The mass bound obtained in this paper is comparable to our result in the previous paragraph.}. 
For example, the cross section of the process $e^+ e^- \to E_1^+ E_1^-$ is calculated to be 
\begin{equation}
  \sigma(e^+ e^- \to E_1^+ E_1^-) = \frac{y_\varphi^4}{32 \pi s} 
  \left( \beta_{KK} + \frac{m_{\rm KK}^2}{s} \log\left[\frac{s(1+\beta_{KK})-2 m_{\rm KK}^2}{s(1-\beta_{KK})
  -2 m_{\rm KK}^2} \right] 
  \right), 
\end{equation}
where $\beta_{KK} = \sqrt{1-4 m_{\rm KK}^2/s}$. 
We may compare it with the QED production process mediated by photon in the $s$-channel, 
   $e^+ e^- \to \gamma^* \to E_1^+ E_1^-$, whose cross section is given by 
\bea
  \sigma(e^+ e^- \to \gamma^* \to E_1^+ E_1^-) = \frac{e^4}{12 \pi s} \beta_{KK}  \left( 1- 2 \frac{m_{\rm KK}^2}{s} \right),  
\eea 
where $e$ is the QED coupling. 
Hence, unless the collider energy far exceeds the KK-mode mass, 
   the NG boson-mediated process remains negligible for $y_\varphi$ smaller than the SM gauge couplings.

%%%%%%%%%%%%%%%%%%%%%%%%%%%%%%%%%%%%%%%%%%%%%%%%%%%%%%%%
\section{Conclusion and discussion}
\label{sec:4}
%%%%%%%%%%%%%%%%%%%%%%%%%%%%%%%%%%%%%%%%%%%%%%%%%%%%%%%%

In this paper, we have investigated the Domain-Wall Standard Model formulated in five-dimensional spacetime.
In this framework, the SM gauge bosons, Higgs doublet, and chiral fermions are localized on a domain wall 
   along the non-compact extra spatial dimension.
By integrating out the extra-dimensional degrees of freedom, we obtain the SM as an effective four-dimensional theory,
   where SM particles are accompanied by their KK-mode partners.
We have focused on the fermion sector of the Domain-Wall SM, where a NG boson arises 
   due to the spontaneous breaking of translational invariance in the extra dimension. 
The presence of such an NG boson is a distinctive feature of this model, 
   in contrast to conventional braneworld scenarios,
   where extra dimensions are compactified on manifolds or orbifolds with finite volume, 
   explicitly breaking translational invariance.

We have studied phenomenological constraints on the Domain-Wall SM involving the NG boson through the Yukawa couplings among the NG boson, SM fermions, and their KK partners. 
First, we examined the potential energy loss in the supernova SN 1987A via NG boson emission from its core.
We found that if the KK-mode electron mass is significantly higher than the electron chemical potential in the supernova core,
     the scenario remains consistent with the neutrino observations from SN 1987A by the Kamiokande II and IMB collaborations.
Next, we analyzed the thermal history of the early universe.
Due to its couplings, the NG boson stays in thermal equilibrium with the SM plasma in the early universe.
We derived constraints from the additional contribution of NG bosons to the energy density during BBN,
   leading to a bound on the Yukawa coupling and KK-mode fermion masses.
Finally, we discussed the direct detection of KK-mode fermions at the LHC.
The production and decay signatures of KK-mode quarks and leptons closely resemble those of squarks and sleptons,
   apart from differences in particle spins and the presence of a massless NG boson.
By reinterpreting the existing LHC search results for squarks and sleptons in simplified supersymmetric models with a massless lightest neutralino,
   we obtained a conservative lower bound of around 1 TeV for KK-mode quarks (top and bottom) and KK-mode charged leptons.

%%%%%%%%  Fig 2 %%%%%%%%%%%%%%%%%%%%%%%%%%%%%%%
\begin{figure}[t]
\begin{center}
\includegraphics[width=0.7\textwidth]{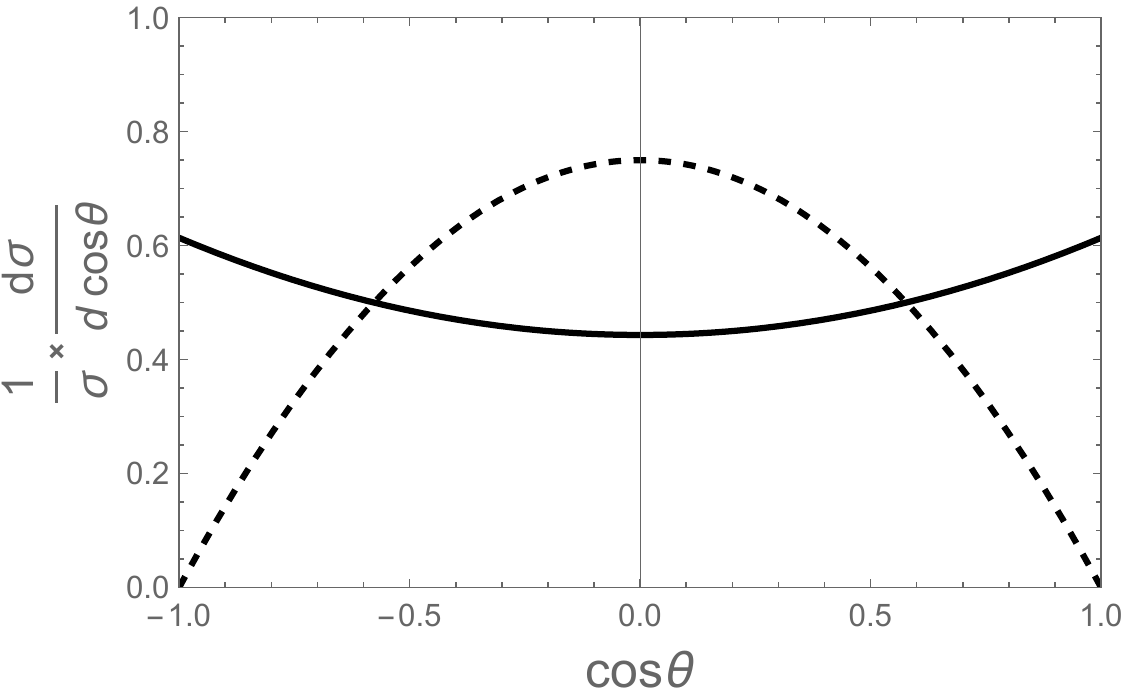} 
\caption{
Normalized differential cross sections for pair productions of KK-mode electrons (black solid)
  and right-handed selectrons (black dashed), respectively.
Here, we set the collider energy $\sqrt{s}=3$ TeV and $m_{E}=m_{{\tilde e}_R}=1$ TeV.
} 
\label{fig:2}
\end{center}
\end{figure}
%%%%%%%%%%%%%%%%

If the evidence of squark or slepton production emerges in future LHC data,
  the events could alternatively be reinterpreted as KK-mode quark and/or lepton production, 
  especially if a massless (or very light) neutralino adequately accounts for the missing energy distribution. 
The difference in particle spins between KK-mode fermions and scalar superpartners manifests 
  in distinct angular distributions of the produced particles.
Proposed future high-energy lepton colliders, such as the International $e^+ e^-$ Linear Collider \cite{Behnke:2013xla}
  and a multi-TeV $\mu^+ \mu^-$ collider \cite{InternationalMuonCollider:2025sys},
  can distinguish between these scenarios by precisely measuring angular distributions of the final states.
To illustrate this, we calculate the differential cross sections for pair production of SU(2)$_L$ singlet KK-mode electrons $(E)$
  and right-handed selectrons $(\tilde{e}_R)$ at a $\mu^+ \mu^-$ collider with $\sqrt{s}=3$ TeV. 
Given the high collider energy, we can neglect weak gauge boson masses and 
  consider massless U(1) hypercharge gauge boson-mediated processes:
\bea
&&\frac{d \sigma}{d \cos\theta}(\mu^+ \mu^- \to \gamma^*, Z^* \to \bar{E} \, E) 
 \simeq  \frac{5}{512 \pi \, s} \, g_Y^4 \, \beta_{E}  \, (4-\beta_{E}^2+\beta_{E}^2 \cos(2 \theta)) , \nonumber\\
&&\frac{d \sigma}{d \cos\theta}(\mu^+ \mu^- \to \gamma^*, Z^* \to {\tilde e}_R^{~*}  \, {\tilde e}_R) 
 \simeq  \frac{5}{512 \pi \, s} \, g_Y^4 \, \beta_{\tilde e}^3  \,  \sin^2 \theta , 
\eea
where $g_Y$ is the U(1) hypercharge gauge coupling, 
  $\beta_{E}=\sqrt{1-4 m_{E}^2/s}$, and $\beta_{\tilde e}=\sqrt{1-4 m_{{\tilde e}_R}^2/s}$. 
Fig.~\ref{fig:2} shows the normalized differential cross sections for these two processes,
  revealing a clear difference in their angular dependence.

Since the Yukawa coupling of the NG boson is a free parameter in our scenario,
    KK-mode fermions can become long-lived if the coupling is sufficiently small.
Long-lived KK-mode quarks form exotic heavy charged hadrons, known as R-hadrons.
Similarly, long-lived KK-mode leptons behave like heavy exotic muons.
Such long-lived charged particles leave highly ionizing tracks in detectors due to their slow traversal.
The ATLAS and CMS collaborations \cite{ATLAS:2019gqq, CMS:2024nhn} have searched for such particles 
    by using observables like ionization energy loss and time of flight.
Their analyses considered long-lived sbottoms, stops, and staus in the Minimal Supersymmetric Standard Model,
   setting lower mass bounds of $1.2-1.4$ TeV for sbottoms and stops, and around 430 GeV for staus.
As previously discussed, the production mechanisms for sbottoms, stops, and staus are essentially identical
   to those of KK-mode bottom quarks, top quarks, and taus,
   though fermion pair production cross sections are typically a few times larger than those of scalar pairs.
Thus, we conservatively adopt these experimental lower bounds as constraints on long-lived KK-mode fermions.

%%%%%%%%%%%%%%%%%%%%%%%%%%%%%%%%%%
\section*{Acknowledgments}
%%%%%%%%%%%%%%%%%%%%%%%%%%%%%%%%%%
This work is supported in part by the United States Department of Energy Grants DE-SC0012447 (P.D. and N.O.) and DE-SC0023713 (N.O.).

\newpage
%%%%%%%%%%%%%%%%%%%%%%%%%%%  
\bibliographystyle{utphysII}
\bibliography{References}

@article{Polchinski:1995mt,
    author = "Polchinski, Joseph",
    title = "{Dirichlet Branes and Ramond-Ramond charges}",
    eprint = "hep-th/9510017",
    archivePrefix = "arXiv",
    reportNumber = "NSF-ITP-95-122",
    doi = "10.1103/PhysRevLett.75.4724",
    journal = "Phys. Rev. Lett.",
    volume = "75",
    pages = "4724--4727",
    year = "1995"
}

@article{Arkani-Hamed:1998jmv,
    author = "Arkani-Hamed, Nima and Dimopoulos, Savas and Dvali, G. R.",
    title = "{The Hierarchy problem and new dimensions at a millimeter}",
    eprint = "hep-ph/9803315",
    archivePrefix = "arXiv",
    reportNumber = "SLAC-PUB-7769, SU-ITP-98-13",
    doi = "10.1016/S0370-2693(98)00466-3",
    journal = "Phys. Lett. B",
    volume = "429",
    pages = "263--272",
    year = "1998"
}

@article{Antoniadis:1998ig,
    author = "Antoniadis, Ignatios and Arkani-Hamed, Nima and Dimopoulos, Savas and Dvali, G. R.",
    title = "{New dimensions at a millimeter to a Fermi and superstrings at a TeV}",
    eprint = "hep-ph/9804398",
    archivePrefix = "arXiv",
    reportNumber = "SLAC-PUB-7801, SU-ITP-98-28, CPTH-S608-0498, IC-98-39",
    doi = "10.1016/S0370-2693(98)00860-0",
    journal = "Phys. Lett. B",
    volume = "436",
    pages = "257--263",
    year = "1998"
}

@article{Randall:1999ee,
    author = "Randall, Lisa and Sundrum, Raman",
    title = "{A Large mass hierarchy from a small extra dimension}",
    eprint = "hep-ph/9905221",
    archivePrefix = "arXiv",
    reportNumber = "MIT-CTP-2860, PUPT-1860, BUHEP-99-9",
    doi = "10.1103/PhysRevLett.83.3370",
    journal = "Phys. Rev. Lett.",
    volume = "83",
    pages = "3370--3373",
    year = "1999"
}

@article{Randall:1999vf,
    author = "Randall, Lisa and Sundrum, Raman",
    title = "{An Alternative to compactification}",
    eprint = "hep-th/9906064",
    archivePrefix = "arXiv",
    reportNumber = "MIT-CTP-2874, PUPT-1867, BUHEP-99-13",
    doi = "10.1103/PhysRevLett.83.4690",
    journal = "Phys. Rev. Lett.",
    volume = "83",
    pages = "4690--4693",
    year = "1999"
}

@article{Binetruy:1999ut,
    author = "Binetruy, Pierre and Deffayet, Cedric and Langlois, David",
    title = "{Nonconventional cosmology from a brane universe}",
    eprint = "hep-th/9905012",
    archivePrefix = "arXiv",
    reportNumber = "LPT-ORSAY-99-25",
    doi = "10.1016/S0550-3213(99)00696-3",
    journal = "Nucl. Phys. B",
    volume = "565",
    pages = "269--287",
    year = "2000"
}

@article{Binetruy:1999hy,
    author = "Binetruy, Pierre and Deffayet, Cedric and Ellwanger, Ulrich and Langlois, David",
    title = "{Brane cosmological evolution in a bulk with cosmological constant}",
    eprint = "hep-th/9910219",
    archivePrefix = "arXiv",
    reportNumber = "LPT-ORSAY-99-87",
    doi = "10.1016/S0370-2693(00)00204-5",
    journal = "Phys. Lett. B",
    volume = "477",
    pages = "285--291",
    year = "2000"
}

@article{Shiromizu:1999wj,
    author = "Shiromizu, Tetsuya and Maeda, Kei-ichi and Sasaki, Misao",
    title = "{The Einstein equation on the 3-brane world}",
    eprint = "gr-qc/9910076",
    archivePrefix = "arXiv",
    reportNumber = "DAMTP-1999-150, OUTAP-103, UTAP-349, RESCEU-40-99",
    doi = "10.1103/PhysRevD.62.024012",
    journal = "Phys. Rev. D",
    volume = "62",
    pages = "024012",
    year = "2000"
}

@article{Ida:1999ui,
    author = "Ida, Daisuke",
    title = "{Brane world cosmology}",
    eprint = "gr-qc/9912002",
    archivePrefix = "arXiv",
    doi = "10.1088/1126-6708/2000/09/014",
    journal = "JHEP",
    volume = "09",
    pages = "014",
    year = "2000"
}

@article{Okada:2017omx,
    author = "Okada, Nobuchika and Raut, Digesh and Villalba, Desmond",
    title = "{Domain-Wall Standard Model in non-compact 5D and LHC phenomenology}",
    eprint = "1712.09323",
    archivePrefix = "arXiv",
    primaryClass = "hep-ph",
    doi = "10.1142/S0217732319500809",
    journal = "Mod. Phys. Lett. A",
    volume = "34",
    number = "10",
    pages = "1950080",
    year = "2019"
}

@article{Okada:2018von,
    author = "Okada, Nobuchika and Raut, Digesh and Villalba, Desmond",
    title = "{Aspects of the Domain-Wall Standard Model}",
    eprint = "1801.03007",
    archivePrefix = "arXiv",
    primaryClass = "hep-ph",
    doi = "10.1093/ptep/ptad155",
    journal = "PTEP",
    volume = "2024",
    number = "2",
    pages = "023B01",
    year = "2024"
}

@article{Arai:2018uoy,
    author = "Arai, Masato and Blaschke, Filip and Eto, Minoru and Sakai, Norisuke",
    title = "{Localization of the Standard Model via the Higgs mechanism and a finite electroweak monopole from non-compact five dimensions}",
    eprint = "1802.06649",
    archivePrefix = "arXiv",
    primaryClass = "hep-ph",
    reportNumber = "YGHP-18-03",
    doi = "10.1093/ptep/pty083",
    journal = "PTEP",
    volume = "2018",
    number = "8",
    pages = "083B04",
    year = "2018"
}

@article{Okada:2019fgm,
    author = "Okada, Nobuchika and Raut, Digesh and Villalba, Desmond",
    title = "{Fermion Mass Hierarchy and Phenomenology in the 5D Domain Wall Standard Model}",
    eprint = "1904.10308",
    archivePrefix = "arXiv",
    primaryClass = "hep-ph",
    doi = "10.1007/JHEP10(2019)259",
    journal = "JHEP",
    volume = "10",
    pages = "259",
    year = "2019"
}

@article{Davies:2007xr,
    author = "Davies, Rhys and George, Damien P. and Volkas, Raymond R.",
    title = "{The Standard model on a domain-wall brane}",
    eprint = "0705.1584",
    archivePrefix = "arXiv",
    primaryClass = "hep-ph",
    doi = "10.1103/PhysRevD.77.124038",
    journal = "Phys. Rev. D",
    volume = "77",
    pages = "124038",
    year = "2008"
}

@article{Arai:2012cx,
    author = "Arai, Masato and Blaschke, Filip and Eto, Minoru and Sakai, Norisuke",
    title = "{Matter Fields and Non-Abelian Gauge Fields Localized on Walls}",
    eprint = "1208.6219",
    archivePrefix = "arXiv",
    primaryClass = "hep-th",
    reportNumber = "YGHP-12-50",
    doi = "10.1093/ptep/pts050",
    journal = "PTEP",
    volume = "2013",
    pages = "013B05",
    year = "2013"
}

@article{Arai:2013mwa,
    author = "Arai, Masato and Blaschke, Filip and Eto, Minoru and Sakai, Norisuke",
    title = "{Stabilizing matter and gauge fields localized on walls}",
    eprint = "1303.5212",
    archivePrefix = "arXiv",
    primaryClass = "hep-th",
    reportNumber = "YGHP-13-56",
    doi = "10.1093/ptep/ptt064",
    journal = "PTEP",
    volume = "2013",
    number = "9",
    pages = "093B01",
    year = "2013"
}

@article{Arai:2017ntb,
    author = "Arai, Masato and Blaschke, Filip and Eto, Minoru and Sakai, Norisuke",
    title = "{Non-Abelian Gauge Field Localization on Walls and Geometric Higgs Mechanism}",
    eprint = "1703.00427",
    archivePrefix = "arXiv",
    primaryClass = "hep-th",
    reportNumber = "YGHP-17-05",
    doi = "10.1093/ptep/ptx047",
    journal = "PTEP",
    volume = "2017",
    number = "5",
    pages = "053B01",
    year = "2017"
}

@article{Arai:2017lfv,
    author = "Arai, Masato and Blaschke, Filip and Eto, Minoru and Sakai, Norisuke",
    title = "{Grand Unified Brane World Scenario}",
    eprint = "1703.00351",
    archivePrefix = "arXiv",
    primaryClass = "hep-th",
    reportNumber = "YGHP-17-06",
    doi = "10.1103/PhysRevD.96.115033",
    journal = "Phys. Rev. D",
    volume = "96",
    number = "11",
    pages = "115033",
    year = "2017"
}

@article{Arai:2021wul,
    author = "Arai, Masato and Blaschke, Filip and Eto, Minoru and Kawaguchi, Masaki and Sakai, Norisuke",
    title = "{Standard model gauge fields localized on non-Abelian vortices in six dimensions}",
    eprint = "2105.06026",
    archivePrefix = "arXiv",
    primaryClass = "hep-th",
    reportNumber = "YGHP-21-2",
    doi = "10.1093/ptep/ptab144",
    journal = "PTEP",
    volume = "2021",
    number = "12",
    pages = "123B07",
    year = "2021"
}

@article{Ohta:2010fu,
    author = "Ohta, Kazutoshi and Sakai, Norisuke",
    title = "{Non-Abelian Gauge Field Localized on Walls with Four-Dimensional World Volume}",
    eprint = "1004.4078",
    archivePrefix = "arXiv",
    primaryClass = "hep-th",
    doi = "10.1143/PTP.124.71",
    journal = "Prog. Theor. Phys.",
    volume = "124",
    pages = "71--93",
    year = "2010",
    note = "[Erratum: Prog.Theor.Phys. 127, 1133 (2012)]"
}

@article{Luty:2002hj,
    author = "Luty, Markus A. and Okada, Nobuchika",
    title = "{Almost no scale supergravity}",
    eprint = "hep-th/0209178",
    archivePrefix = "arXiv",
    reportNumber = "KEK-TH-840, UMD-PP-03-011",
    doi = "10.1088/1126-6708/2003/04/050",
    journal = "JHEP",
    volume = "04",
    pages = "050",
    year = "2003"
}

@article{Dvali:1996xe,
    author = "Dvali, G. R. and Shifman, Mikhail A.",
    title = "{Domain walls in strongly coupled theories}",
    eprint = "hep-th/9612128",
    archivePrefix = "arXiv",
    reportNumber = "CERN-TH-96-356, TPI-MINN-96-26-T, UMN-TH-1521-96",
    doi = "10.1016/S0370-2693(97)00131-7",
    journal = "Phys. Lett. B",
    volume = "396",
    pages = "64--69",
    year = "1997",
    note = "[Erratum: Phys.Lett.B 407, 452 (1997)]"
}

@article{Arai:2018rwf,
    author = "Arai, Masato and Blaschke, Filip and Eto, Minoru and Sakai, Norisuke",
    title = "{Localized non-Abelian gauge fields in non-compact extra-dimensions}",
    eprint = "1801.02498",
    archivePrefix = "arXiv",
    primaryClass = "hep-th",
    reportNumber = "YGHP-18-02",
    doi = "10.1093/ptep/pty057",
    journal = "PTEP",
    volume = "2018",
    number = "6",
    pages = "063B02",
    year = "2018"
}

@article{Rubakov:1983bb,
    author = "Rubakov, V. A. and Shaposhnikov, M. E.",
    title = "{Do We Live Inside a Domain Wall?}",
    doi = "10.1016/0370-2693(83)91253-4",
    journal = "Phys. Lett. B",
    volume = "125",
    pages = "136--138",
    year = "1983"
}

@article{Hisano:1999bn,
    author = "Hisano, Junji and Okada, Nobuchika",
    title = "{On effective theory of brane world with small tension}",
    eprint = "hep-ph/9909555",
    archivePrefix = "arXiv",
    reportNumber = "KEK-TH-651",
    doi = "10.1103/PhysRevD.61.106003",
    journal = "Phys. Rev. D",
    volume = "61",
    pages = "106003",
    year = "2000"
}

@article{Kamiokande-II:1987idp,
    author = "Hirata, K. and others",
    editor = "Wali, K. C.",
    collaboration = "Kamiokande-II",
    title = "{Observation of a Neutrino Burst from the Supernova SN 1987a}",
    reportNumber = "UT-ICEPP-87-01, UPR-142E",
    doi = "10.1103/PhysRevLett.58.1490",
    journal = "Phys. Rev. Lett.",
    volume = "58",
    pages = "1490--1493",
    year = "1987"
}

@article{Bionta:1987qt,
    author = "Bionta, R. M. and others",
    title = "{Observation of a Neutrino Burst in Coincidence with Supernova SN 1987a in the Large Magellanic Cloud}",
    reportNumber = "UCI-NEUTRINO-87-10",
    doi = "10.1103/PhysRevLett.58.1494",
    journal = "Phys. Rev. Lett.",
    volume = "58",
    pages = "1494",
    year = "1987"
}

@book{Raffelt:1996wa,
    author = "Raffelt, G. G.",
    title = "{Stars as laboratories for fundamental physics}: {The astrophysics of neutrinos, axions, and other weakly interacting particles}",
    isbn = "978-0-226-70272-8",
    month = "5",
    year = "1996"
}

@article{Burrows:1986me,
    author = "Burrows, Adam and Lattimer, James M.",
    title = "{The birth of neutron stars}",
    doi = "10.1086/164405",
    journal = "Astrophys. J.",
    volume = "307",
    pages = "178--196",
    year = "1986"
}

@article{Planck:2018vyg,
    author = "Aghanim, N. and others",
    collaboration = "Planck",
    title = "{Planck 2018 results. VI. Cosmological parameters}",
    eprint = "1807.06209",
    archivePrefix = "arXiv",
    primaryClass = "astro-ph.CO",
    doi = "10.1051/0004-6361/201833910",
    journal = "Astron. Astrophys.",
    volume = "641",
    pages = "A6",
    year = "2020",
    note = "[Erratum: Astron.Astrophys. 652, C4 (2021)]"
}

@article{Escudero:2018mvt,
    author = "Escudero, Miguel",
    title = "{Neutrino decoupling beyond the Standard Model: CMB constraints on the Dark Matter mass with a fast and precise $N_{\rm eff}$ evaluation}",
    eprint = "1812.05605",
    archivePrefix = "arXiv",
    primaryClass = "hep-ph",
    reportNumber = "KCL-2018-76",
    doi = "10.1088/1475-7516/2019/02/007",
    journal = "JCAP",
    volume = "02",
    pages = "007",
    year = "2019"
}

@article{Bennett:2019ewm,
    author = "Bennett, Jack J. and Buldgen, Gilles and Drewes, Marco and Wong, Yvonne Y. Y.",
    title = "{Towards a precision calculation of the effective number of neutrinos $N_{\rm eff}$ in the Standard Model I: the QED equation of state}",
    eprint = "1911.04504",
    archivePrefix = "arXiv",
    primaryClass = "hep-ph",
    doi = "10.1088/1475-7516/2020/03/003",
    journal = "JCAP",
    volume = "03",
    pages = "003",
    year = "2020",
    note = "[Addendum: JCAP 03, A01 (2021)]"
}

@article{Escudero:2020dfa,
    author = "Escudero Abenza, Miguel",
    title = "{Precision early universe thermodynamics made simple: $N_{\rm eff}$ and neutrino decoupling in the Standard Model and beyond}",
    eprint = "2001.04466",
    archivePrefix = "arXiv",
    primaryClass = "hep-ph",
    reportNumber = "KCL-2019-85",
    doi = "10.1088/1475-7516/2020/05/048",
    journal = "JCAP",
    volume = "05",
    pages = "048",
    year = "2020"
}

@article{Akita:2020szl,
    author = "Akita, Kensuke and Yamaguchi, Masahide",
    title = "{A precision calculation of relic neutrino decoupling}",
    eprint = "2005.07047",
    archivePrefix = "arXiv",
    primaryClass = "hep-ph",
    doi = "10.1088/1475-7516/2020/08/012",
    journal = "JCAP",
    volume = "08",
    pages = "012",
    year = "2020"
}

@article{Bennett:2020zkv,
    author = "Bennett, Jack J. and Buldgen, Gilles and De Salas, Pablo F. and Drewes, Marco and Gariazzo, Stefano and Pastor, Sergio and Wong, Yvonne Y. Y.",
    title = "{Towards a precision calculation of $N_{\rm eff}$ in the Standard Model II: Neutrino decoupling in the presence of flavour oscillations and finite-temperature QED}",
    eprint = "2012.02726",
    archivePrefix = "arXiv",
    primaryClass = "hep-ph",
    reportNumber = "CPPC-2020-10",
    doi = "10.1088/1475-7516/2021/04/073",
    journal = "JCAP",
    volume = "04",
    pages = "073",
    year = "2021"
}

@article{Alwall:2008va,
    author = "Alwall, Johan and Le, My-Phuong and Lisanti, Mariangela and Wacker, Jay G.",
    title = "{Model-Independent Jets plus Missing Energy Searches}",
    eprint = "0809.3264",
    archivePrefix = "arXiv",
    primaryClass = "hep-ph",
    reportNumber = "SLAC-PUB-13405",
    doi = "10.1103/PhysRevD.79.015005",
    journal = "Phys. Rev. D",
    volume = "79",
    pages = "015005",
    year = "2009"
}

@article{Alwall:2008ag,
    author = "Alwall, Johan and Schuster, Philip and Toro, Natalia",
    title = "{Simplified Models for a First Characterization of New Physics at the LHC}",
    eprint = "0810.3921",
    archivePrefix = "arXiv",
    primaryClass = "hep-ph",
    reportNumber = "SLAC-PUB-13425, SU-ITP-08-24",
    doi = "10.1103/PhysRevD.79.075020",
    journal = "Phys. Rev. D",
    volume = "79",
    pages = "075020",
    year = "2009"
}

@article{LHCNewPhysicsWorkingGroup:2011mji,
    author = "Alves, Daniele",
    editor = "Arkani-Hamed, Nima and others",
    collaboration = "LHC New Physics Working Group",
    title = "{Simplified Models for LHC New Physics Searches}",
    eprint = "1105.2838",
    archivePrefix = "arXiv",
    primaryClass = "hep-ph",
    reportNumber = "SLAC-PUB-15045, FERMILAB-PUB-11-842-A-PPD",
    doi = "10.1088/0954-3899/39/10/105005",
    journal = "J. Phys. G",
    volume = "39",
    pages = "105005",
    year = "2012"
}

@article{ATLAS:2024lda,
    author = "Aad, Georges and others",
    collaboration = "ATLAS",
    title = "{The quest to discover supersymmetry at the ATLAS experiment}",
    eprint = "2403.02455",
    archivePrefix = "arXiv",
    primaryClass = "hep-ex",
    reportNumber = "CERN-EP-2024-056",
    doi = "10.1016/j.physrep.2024.09.010",
    journal = "Phys. Rept.",
    volume = "1116",
    pages = "261--300",
    year = "2025"
}

@article{Behnke:2013xla,
    editor = "Behnke, Ties and Brau, James E. and Foster, Brian and Fuster, Juan and Harrison, Mike and Paterson, James McEwan and Peskin, Michael and Stanitzki, Marcel and Walker, Nicholas and Yamamoto, Hitoshi",
    title = "{The International Linear Collider Technical Design Report - Volume 1: Executive Summary}",
    eprint = "1306.6327",
    archivePrefix = "arXiv",
    primaryClass = "physics.acc-ph",
    reportNumber = "ILC-REPORT-2013-040, ANL-HEP-TR-13-20, BNL-100603-2013-IR, IRFU-13-59, CERN-ATS-2013-037, COCKCROFT-13-10, CLNS-13-2085, DESY-13-062, FERMILAB-TM-2554, IHEP-AC-ILC-2013-001, INFN-13-04-LNF, JAI-2013-001, JINR-E9-2013-35, JLAB-R-2013-01, KEK-REPORT-2013-1, KNU-CHEP-ILC-2013-1, LLNL-TR-635539, SLAC-R-1004, ILC-HIGRADE-REPORT-2013-003",
    month = "6",
    year = "2013"
}

@article{InternationalMuonCollider:2025sys,
    author = "Accettura, Carlotta and others",
    collaboration = "International Muon Collider",
    title = "{The Muon Collider}",
    eprint = "2504.21417",
    archivePrefix = "arXiv",
    primaryClass = "physics.acc-ph",
    reportNumber = "FERMILAB-PUB-25-0309-AD-PPD-T",
    month = "4",
    year = "2025"
}

@article{ATLAS:2019gqq,
    author = "Aaboud, Morad and others",
    collaboration = "ATLAS",
    title = "{Search for heavy charged long-lived particles in the ATLAS detector in 36.1 fb$^{-1}$ of proton-proton collision data at $\sqrt{s} = 13$ TeV}",
    eprint = "1902.01636",
    archivePrefix = "arXiv",
    primaryClass = "hep-ex",
    reportNumber = "CERN-EP-2018-339",
    doi = "10.1103/PhysRevD.99.092007",
    journal = "Phys. Rev. D",
    volume = "99",
    number = "9",
    pages = "092007",
    year = "2019"
}

@article{CMS:2024nhn,
    author = "Hayrapetyan, Aram and others",
    collaboration = "CMS",
    title = "{Search for heavy long-lived charged particles with large ionization energy loss in proton-proton collisions at $ \sqrt{s} $ = 13 TeV}",
    eprint = "2410.09164",
    archivePrefix = "arXiv",
    primaryClass = "hep-ex",
    reportNumber = "CMS-EXO-18-002, CERN-EP-2024-254",
    doi = "10.1007/JHEP04(2025)109",
    journal = "JHEP",
    volume = "04",
    pages = "109",
    year = "2025"
}

\end{document}